\documentclass[prl,twocolumn,english,showpacs]{revtex4-1}
\usepackage{graphicx}
\usepackage{amssymb}

\begin{document}

\title{A single ion coupled to an optical fiber cavity}

\author{Matthias Steiner$^{1}$, Hendrik M. Meyer$^{1}$, Christian Deutsch$^{2,3}$, Jakob Reichel$^2$, and Michael K{\"o}hl$^1$}

\affiliation{$^1${Cavendish Laboratory, University of Cambridge, JJ Thomson Avenue, Cambridge CB3 0HE, UK}\\$^2${Laboratoire Kastler-Brossel, ENS/UPMC-Paris 6/CNRS, F-75005 Paris, France}\\$^3$Menlo Systems GmbH, 82152 Martinsried, Germany}

\begin{abstract}
We present the realization of a combined trapped-ion and optical cavity system, in which a single Yb$^+$ ion is confined by a micron-scale ion trap inside a 230\,$\mu$m-long optical fiber cavity. We characterize the spatial ion-cavity coupling and measure the ion-cavity coupling strength using a cavity-stimulated $\Lambda$-transition. Owing to the small mode volume of the fiber resonator, the coherent coupling strength between the ion and a single photon exceeds the natural decay rate of the dipole moment. This system can be integrated into ion-photon quantum networks and is a step towards cavity quantum-electrodynamics (cavity-QED) based quantum information processing with trapped ions.
\end{abstract}

\date{\today}

\maketitle

Trapped atomic ions play an important role in studies of small, isolated quantum systems, for example in quantum information processing and precision metrology. Aside from the long achievable coherence times, their success is largely based on the excellent manipulation and interrogation possibilities of their internal quantum states, which are usually performed by optical means. In order to employ the outstanding properties of trapped ions for future applications such as cavity-QED based quantum computers \cite{Pellizzari1995} or quantum network nodes \cite{Cirac1997,Kimble2008,Duan2010}, strong coupling between a single ion and a single photon is a prerequisite, i.e., the coherent coupling strength must exceed the decoherence rate of the atomic dipole moment. Unlike for neutral atoms \cite{Brune1996,Boca2004} and solid-state emitters, such as quantum dots \cite{Reithmaier2004} or Cooper pairs \cite{Wallraff2004}, this strong coupling regime has not yet been reached for a single trapped ion despite decade-long efforts \cite{Guthoerlein2001,Mundt2002,Keller2004,Leibrandt2009,Herskind2009,Sterk2011,Stute2012}.

The route to achieve strong light-matter coupling employs the principles of cavity-QED; a resonator changes the mode structure of the vacuum electromagnetic field in order to strongly enhance coupling to one photon mode. The coupling strength $g$ between a single emitter and a single-photon mode depends on the mode-volume $V$ of the cavity and on the electric dipole moment $d$ of the transition, $g \propto d/\sqrt{V}$. Owing to the large mode volumes of the cavities used in previous experiments \cite{Guthoerlein2001,Mundt2002,Keller2004,Leibrandt2009,Herskind2009,Sterk2011,Stute2012}, the coherent single-photon coupling rate $g$ has been inferior to the decay rate $\Gamma$ of the atomic dipole moment. The main restriction has been that the crucial ingredient to achieve strong coupling, namely placing the ion near dielectric mirror surfaces which are necessary to form an optical cavity, has been found to severely compromise the performance of a Paul trap \cite{Harlander2010}. For sizing down both the mode-volume and the amount of dielectric material, the development of cavities based on optical fibers \cite{Hunger2010} has opened a new perspective, also with respect to the integration of optical elements into microchip-based ion traps. Optical fiber cavities offer significantly smaller radii of curvature of the mirrors, which lead to a small waist of the field mode inside the optical cavity. Recently, significant experimental efforts have been devoted to integrating optical fibers with ion traps for efficient light collection \cite{Vandevender2010,Herskind2011,Wilson2011,Kim2011} but optical cavities have not yet been realized.

Here, we demonstrate coupling of a single trapped Yb$^+$ ion to a microscopic optical cavity. The cavity is formed by a pair of micromachined and reflectively coated ends of optical fibers \cite{Hunger2010}, which provide a small dielectric surface area and a small radius of curvature for reducing the mode volume. The cavity has a length of $L=230$\,$\mu$m and a mode waist of $w=7$\,$\mu$m, leading to an overall reduction of the mode volume by more than two orders of magnitude as compared to previous work. We observe the spatially dependent coupling of the ion to the field of the cavity and measure the coherent ion-cavity coupling strength to be larger than the decay rate of the atomic dipole moment.

\begin{figure}
 \includegraphics[width=\columnwidth,clip=true]{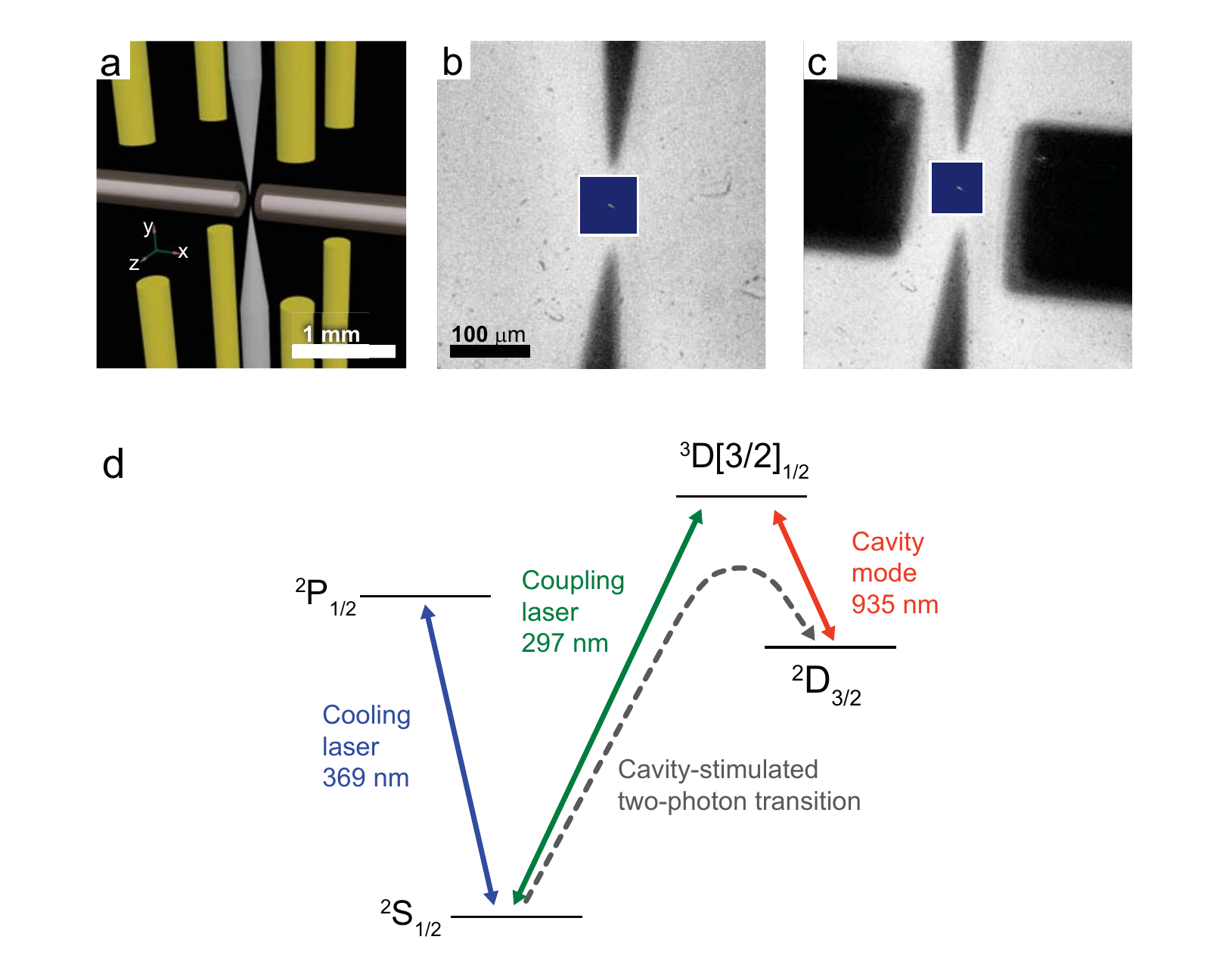}
 \caption{(Color online) Experimental setup and relevant energy levels of Yb$^+$. \textbf{a} The ion trap is formed by endcap electrodes (grey, vertical), which are surrounded by compensation electrodes (yellow, arranged on a square) and the optical cavity (oriented horizontally). For loading and initial cooling of the ion, the cavity is retracted using a nanopositioning stage. \textbf{b} Composite image of the endcap electrodes (shadow image) and the trapped ion (fluorescence image, inset). \textbf{c} Composite image with the cavity in position. \textbf{d} Cooling and detection of the ion is performed on the $S_{1/2}-\,P_{1/2}$ transition, the cavity operates on the $D_{3/2}-\,D[3/2]_{1/2}$ transition at 935\,nm, and the $\Lambda$-transition is driven by a coupling laser at 297\,nm and the vacuum cavity-mode.}
 \label{fig1}
\end{figure}

Our principal experimental setup is shown in Figure 1a. The radiofrequency (r.f.) Paul trap consists of two opposing endcap electrodes \cite{Schrama1993} with an ion-electrode separation of 50 $\mu$m. The electrodes are made of tungsten wire etched to a cone-like geometry \cite{Deslauriers2006,Meyer2012} and are powered by an r.f. signal of 22\,MHz frequency and 160\,Volts amplitude. Each endcap electrode is surrounded by four compensation electrodes formed by gold wires of 250\,$\mu$m diameter arranged on a square of side length 1.2\,mm and recessed by 500\,$\mu$m, which constitute r.f. ground and which are employed for excess micromotion compensation (see Figure 1a). We load single Yb$^+$ ions by Doppler-free photoionization from a pulsed atomic beam \cite{Zipkes2010} and, subsequently, we laser-cool the ion on the $S_{1/2}-P_{1/2}$ transition near 369\,nm. During laser cooling, spontaneous decay from the excited $P_{1/2}$ state populates the metastable $D_{3/2}$ state, which is cleared out by a repumping laser at 935\,nm. The secular trap frequencies are $\omega_{x}/2\pi= 2.5$~MHz and $\omega_{y}/2\pi=\omega_{z}/2\pi= 1.5$~MHz. We have measured the heating rate of the trapped ion using the Doppler-recooling method \cite{Wesenberg2007} and find $(1.0\pm 0.1)$~K/s, in line with previous measurements of similar trap geometries \cite{Deslauriers2006}.

The fiber cavity is aligned in the plane orthogonal to the symmetry axis of the ion trap. We mount the fiber ends in stainless steel sleeves with an outer diameter of 300\,$\mu$m in order to provide mechanical stability, electrical grounding, and shielding of the fibers from  ultraviolet stray light, which would charge up the dielectric material. During loading and initial laser cooling of the ion, the fiber cavity is located 1.8\,mm away from the center of the ion trap in order to prevent contamination of the mirrors with neutral Yb atoms from the atomic beam. Subsequently, the cavity mode is overlapped with the trapped ion by translating the cavity assembly towards the ion trap using a nanopositioning stage with a speed of 1\,mm/s, during which we maintain laser-cooling of the ion. Experimentally, we find that the cold ion remains trapped during the overlap procedure with the cavity in $>90\%$ of the attempts and that the cavity alignment is unaffected by the transport. We emphasize that our methodology provides an alternative to the concept of having spatially separated loading and processing regions in ion traps \cite{Guthoerlein2001}. When the cavity is in the final position (see Fig. 1c), the length of the optical cavity is actively stabilized by a piezo-electric transducer to a secondary light field near 780\,nm using the Pound-Drever-Hall technique in transmission. The presence of the r.f.-grounded fiber sleeves increases the geometric quadrupole efficiency of the trap from $10\%$ to approximately $19\%$. Accordingly, we reduce the r.f. power and reach secular trap frequencies of $\omega_{x}/2\pi= 2.1$~MHz, $\omega_{y}/2\pi= 2.7$~MHz, and $\omega_{z}/2\pi= 1.3$~MHz. Previously, the heating of trapped ions due to fluctuating potentials on nearby dielectric surfaces has been identified as a critical limitation for integrating ions traps and small optical cavities \cite{Harlander2010}. We find that for our case with $\sim 110\,\mu$m spacing between ion and cavity mirror the ion heating rate doubles to $(2.3\pm 0.4)$~K/s, which is significant but not prohibitively large.

The optical fiber cavity interacts with the Yb$^+$ ion on the $D_{3/2}-\,D[3/2]_{1/2}$ transition near $\lambda=935$\,nm (see Figure 1d). The fiber cavity is designed to provide a peak single-ion/single-photon coupling strength of $g=d \sqrt{4 c/(\hbar \epsilon_0 \lambda w^2 L)}=2 \pi \times 6\,$MHz, where $c$ is the speed of light and $\epsilon_0$ is the vacuum permittivity. The cavity field has a (in-situ) measured decay rate of $\kappa=2 \pi \times 320$\,MHz, corresponding to a finesse of $F=1000\pm50$, which is limited by scattering and absorption losses of the mirrors. Taking into account the branching of the decay of the excited state, the total decay rate of the dipole moment amounts to $\Gamma=2 \pi \times 2$\,MHz and hence $C=0.05$.

\begin{figure*}
 \includegraphics[width=\textwidth,clip=true]{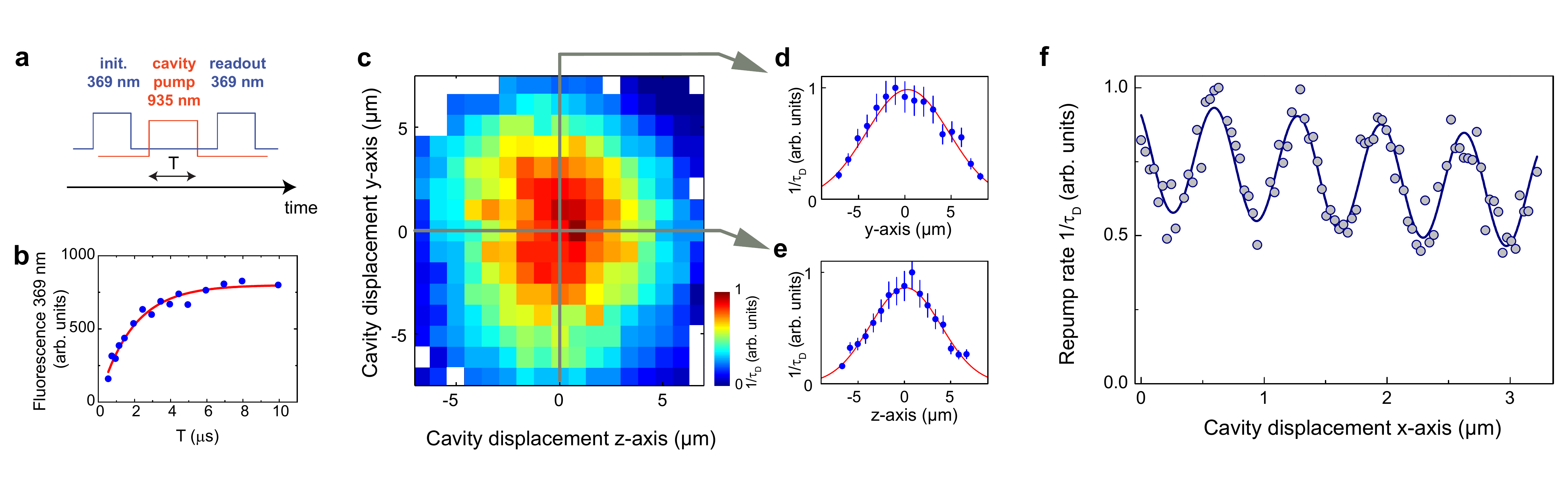}
 \caption{(Color online) Scanning the cavity mode profile. \textbf{a} Laser pulse sequence. An initialization pulse prepares the ion in state $D_{3/2}$ out of which it is repumped using photons on the cavity mode. Subsequently fluorescence on the 369-nm transition is detected. \textbf{b} The repumping rate $1/\tau_D$ out of the $D$-state is determined by an exponential fit $A[1-\exp(-T/\tau_D)]$ to the fluorescence data. \textbf{c} Transverse mode profile showing the intensity profile of the cavity field. \textbf{d} and \textbf{e} show cuts through the mode profile together with Gaussian fits to determine the mode waist. \textbf{f} The optical standing wave inside the resonator is scanned by displacing the cavity assembly relative to the ion trap. The relative error on each data point is $15\%$, given by the fit error to determine $1/\tau_D$.}
 \label{fig2}
\end{figure*}

We demonstrate control over the ion-cavity coupling by displacing the cavity with respect to the ion \cite{Guthoerlein2001}. To this end, we mechanically translate the cavity structure using the nanopositioning stage and, simultaneously, measure the optical pump rate $1/\tau_D$ out of the $D_{3/2}$ state. After preparation of the ion in the $D_{3/2}$ state, we employ laser light on the cavity mode to optically pump into the $S_{1/2}$ ground state, which we detect by fluorescence on the $S_{1/2}-P_{1/2}$ transition (see Figure 2a and 2b). In Figure 2c we display a map of the cross-section of the cavity mode together with Gaussian fits along the $y-$ and the $z-$axes (Figures 2d and 2e). We find waists of $w_y=(7.6\pm0.9)\mu$m and $w_z=(6.6\pm0.9)\mu$m, in agreement with the measured radii of curvature of the mirrors. Owing to the small radius of curvature available with the optical fiber cavity technology, these mode waists are significantly smaller than in previous cavities used with ion traps and employing conventional mirrors \cite{Guthoerlein2001,Mundt2002,Keller2004,Leibrandt2009,Herskind2009,Stute2012}. In Figure 2f we show the standing-wave pattern as the cavity assembly is displaced along the cavity axis. A large repump rate $1/\tau_D$ demonstrates localization of the ion in an antinode of the cavity standing wave. The standing wave pattern shows a period of $(707\pm14)$\,nm, which is larger than half of the  cavity resonant wavelength. The differential shift between the ion and the cavity mode results from the simultaneous displacement of the equilibrium position of the ion as a result of the translation of the r.f.-grounded cavity assembly. It is in agreement with numerical simulations of the ion trap electric fields. We observe a standing-wave contrast of $40\%$ which indicates a localisation of the ion \cite{Guthoerlein2001} to 140\,nm. This localisation could be limited by imperfect Doppler-cooling along the cavity axis, the degree of excess micromotion compensation, and by mechanical vibrations of the cavity assembly relative to the ion trap.

Finally, we demonstrate that our cavity provides coupling between the ion and a single photon of the cavity mode with $g>\Gamma$. In order to measure $g$, we drive a cavity-stimulated $\Lambda-$transition \cite{Hennrich2000} between the $S_{1/2}$ ground state and the $D_{3/2}$ state (see Figure 1d). In this $\Lambda-$scheme, one leg is provided by laser light at 297\,nm, resonant with the $S_{1/2}-D[3/2]_{1/2}$ transition \cite{Meyer2012}, and the other leg is the vacuum cavity mode with coupling strength $g$. We initialize the ion in the $S_{1/2}$ ground state by extinguishing all lasers, followed by a short (5\,$\mu$s) pulse of the 935-nm laser to repump the ion out of the $D_{3/2}$ state into the $S_{1/2}$ state. The  measured repumping time (1/e-time) out of the $D_{3/2}$ state is typically 500\,ns, and therefore the 5\,$\mu$s pulse ensures a full cleanout of the $D_{3/2}$ state \cite{Ratschbacher2012}. Then, we drive the two-photon process and measure its success probability using fluorescence on the $S_{1/2}-P_{1/2}$ transition, as before. In Figure 3 we show the $S_{1/2}$ population as a function of the pumping time $T_{297}$ on the $S_{1/2}-D[3/2]_{1/2}$ transition for both the cavity being on resonance with the $D_{3/2}-D[3/2]_{1/2}$ transition and for being approximately half a free spectral range ($\Delta \nu_{FSR}=650$\,GHz) detuned. We fit the data with an exponentially decaying function $\exp(-T_{297}/\tau)$ and determine the time constants to be $\tau=(34\pm1)\,\mu$s and $\tau=(17.2\pm0.5)\,\mu$s for the case without and with resonant cavity, respectively. To extract the physical parameters from the dissipative system, we model the driven three-level system with a quantized light field numerically using a master equation for the density matrix. We use the case where the cavity is off-resonant, to determine the single-photon Rabi-frequency  $\Omega_{297}=2 \pi \times (1.13\pm0.02)$\,MHz. The solid angle of the cavity mode is only $\delta \Omega=4\times 10^{-4}$ and therefore the off-resonant cavity does not measurably influence the spontaneous emission of the ion. The value of $\Omega_{297}$ is then used in the simulation for the case of a resonant cavity, for which we find $g=2 \pi \times (3.4\pm0.2)$\,MHz, which is larger than $\Gamma$ by almost a factor of two. The coupling strength $g$ is smaller than our peak value of 6\,MHz, but it is in agreement with the results of the standing wave measurement (Figure 2f), which imply a maximum visible coupling strength of $g=2 \pi \times 3.6$\,MHz.

\begin{figure}
 \includegraphics[width=\columnwidth,clip=true]{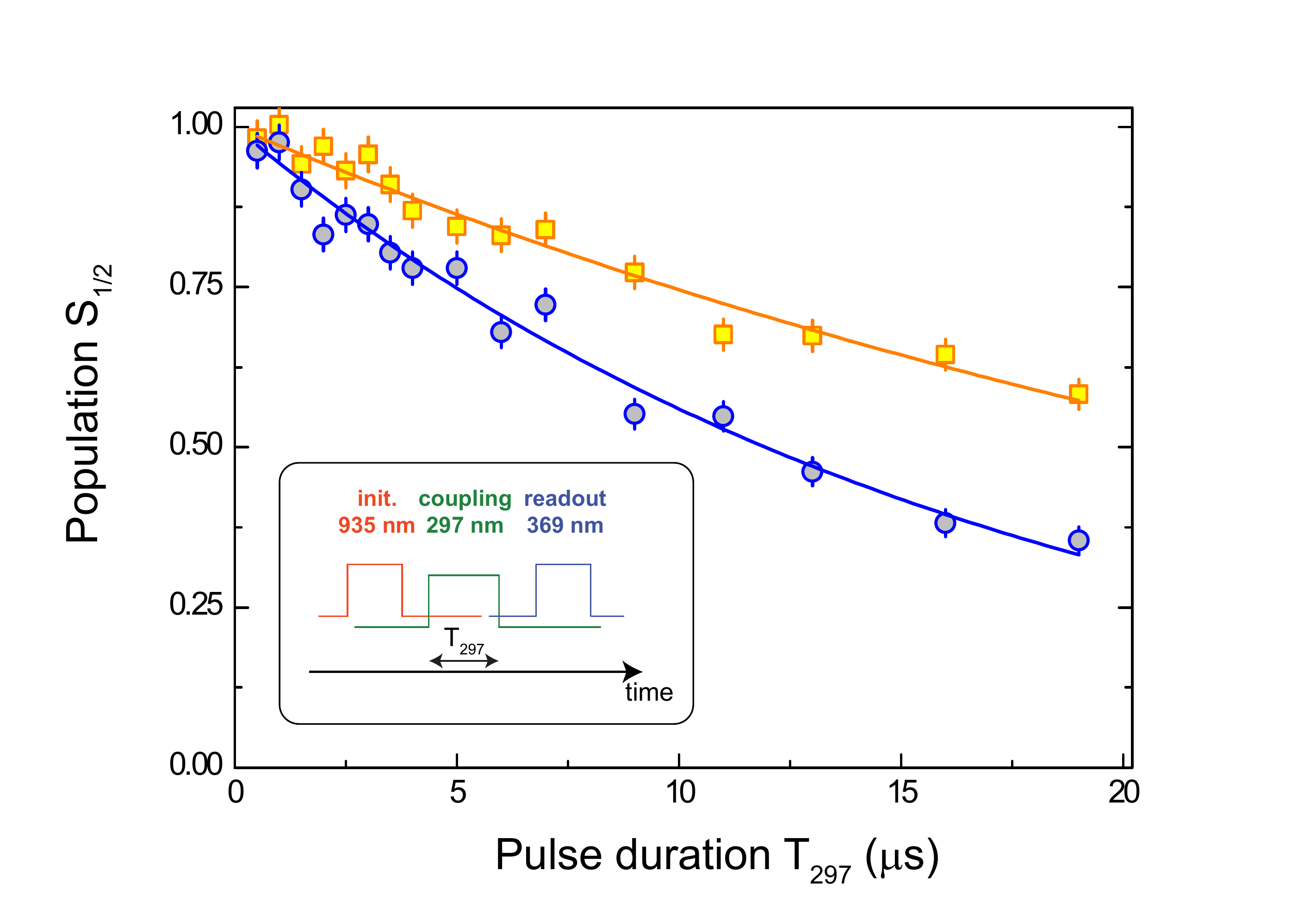}
 \caption{(Color online) Cavity-stimulated two-photon transition rate. Round symbols show the population of the $S_{1/2}$ state after applying a resonant laser on the $S_{1/2}-\,D[3/2]_{1/2}$ and the cavity vacuum field on the $D[3/2]_{1/2}-D_{3/2}$ transitions as a function of the interaction time. The square symbols show the data with the cavity detuned from the atomic transition by half a free spectral range. The solid lines are exponential fits. The inset shows the laser pulse sequence used. The error bars denote the photon shot noise.}
 \label{fig3}
\end{figure}

The large single-ion/single-photon coupling demonstrated here is independent of the cavity finesse but depends crucially on the small cavity mode volume. Our measurements demonstrate implicitly that stable trapping of single ions in close vicinity of dielectric surfaces does not impose fundamental problems, even at room temperature. In the future, using even smaller optical fiber cavities with higher finesse ($F=3\times 10^4$ and $85\%$ coupling between the cavity field and a single mode fiber have been demonstrated for cavities with smaller dimensions \cite{Hunger2010}) will pave the way towards engineering large clusters of remotely entangled single ions in a fiber optical network. Furthermore, in our spectral range the coupling of ions to solid-state emitters, such as semiconductor quantum dots, could be implemented \cite{Waks2009}.

We thank R.T. Phillips and C. Zipkes for support. This work has been supported by EPSRC (EP/H005679/1), ERC (Grant No. 240335), the Royal Society, and the Wolfson Foundation.

\end{document}